\documentclass[prd,aps,twocolumn,a4paper,floatfix,nofootinbib]{revtex4-1}
   
\usepackage[utf8]{inputenc}
\usepackage{graphicx,psfrag}
\usepackage{mathrsfs}
\usepackage{amsmath,amsfonts,amssymb}
\usepackage{multirow}
\usepackage{comment}
\usepackage{xcolor}
\usepackage{enumerate}
\usepackage{hyperref}
\usepackage[T1]{fontenc}
\hypersetup{
    colorlinks = true,
    linkcolor = {blue},
    citecolor = {blue},
    urlcolor = {blue},
    linkbordercolor = {white},
    }

\newcommand{\be}{\begin{equation}}
\newcommand{\ee}{\end{equation}}
\newcommand{\bea}{\begin{eqnarray}}
\newcommand{\eea}{\end{eqnarray}}
\newcommand{\bel}{\begin{align}}
\newcommand{\eel}{\end{align}}

\newcommand{\sone}{S$_{1.25}$ }
\newcommand{\stwo}{S$_{1.50}$ }
\newcommand{\sthree}{S$_{1.75}$ }
\newcommand{\sfour}{S$_{2.00}$ }

\def\i{{\rm i}}

\def\GMc2{{\rm G M_{\odot} c^{-2}}}

\def\z4c{\texttt{Z4c}}

\usepackage{color}
\definecolor{cyan}{rgb}{0,0.9,0.9}
\definecolor{orange}{rgb}{0.9,0.5,0}
\definecolor{magenta}{rgb}{1,0,1}
\definecolor{purple}{rgb}{0.8,0.4,0.8}
\definecolor{gray}{rgb}{0.8242,0.8242,0.8242}
\definecolor{mgreen}{rgb}{0.1,0.8,0.1}

\usepackage[normalem]{ulem}

\begin{document}

\title{High-accuracy high-mass ratio simulations for binary neutron 
       stars and their comparison to existing waveform models}

\author{Maximiliano \surname{Ujevic}$^{1}$} 
\author{Alireza \surname{Rashti}$^{2}$}
\author{Henrique \surname{Gieg}$^{1}$}
\author{Wolfgang \surname{Tichy}$^{2}$}
\author{Tim \surname{Dietrich}$^{3,4}$}

\affiliation{${}^1$ Centro de Ciências Naturais e Humanas, 
Universidade Federal do ABC, 09210-170, Santo Andr{é}, São 
Paulo,Brazil}

\affiliation{${}^2$ Department of Physics, Florida Atlantic 
University, Boca Raton, FL 33431, USA}

\affiliation{${}^3$ Institut für Physik und Astronomie, Universität 
Potsdam, Haus 28, Karl-Liebknecht-Str. 24/25, D-14476, Potsdam, 
Germany}

\affiliation{${}^4$ Max Planck Institute for Gravitational Physics 
(Albert Einstein Institute), Am M{\"u}hlenberg 1, Potsdam 14476, 
Germany}

\date{\today}

\begin{abstract}
The subsequent observing runs of the advanced gravitational-wave detector network will likely provide us with various gravitational-wave observations of binary neutron star systems. For an accurate interpretation of these detections, we need reliable gravitational-wave models. To test and to point out how existing models could be improved, we perform a set of high-resolution numerical-relativity simulations for four different physical setups with mass ratios $q=1.25,~1.50,~1.75,~2.00$, and total gravitational mass $M=2.7M_\odot$. Each configuration is simulated with five different resolutions to allow a proper error assessment. 
Overall, we find approximately 2nd order converging results for the dominant (2,2), but also subdominant (2,1), (3,3), (4,4) modes, while, generally, the convergence order reduces slightly for an increasing mass ratio. Our simulations allow us to validate waveform models, where we find generally good agreement between state-of-the-art models and our data, and to prove that scaling relations for higher modes currently employed for binary black hole waveform modeling also apply for the tidal contribution. Finally, we also test if the current NRTidal model to describe tidal effects is a valid description for high-mass ratio systems. We hope that our simulation results can be used to further improve and test waveform models in preparation for the next observing runs. 
\end{abstract}

\maketitle

\section{Introduction}
\label{sec:intro}

The detection of GW170817~\cite{TheLIGOScientific:2017qsa,LIGOScientific:2017ync,Monitor:2017mdv} and the connected electromagnetic counterparts revolutionized astronomy. This observation of a binary neutron star (BNS) merger led to a variety of important scientific results, e.g., new constraints on the nature of matter at supranuclear densities~\cite{Annala:2017llu,Bauswein:2017vtn,Fattoyev:2017jql,Ruiz:2017due,Shibata:2017xdx,Radice:2017lry,De:2018uhw,Most:2018hfd,Capano:2019eae,Abbott:2018wiz,Abbott:2018exr,Coughlin:2018miv,Coughlin:2018fis,Dietrich:2020lps,Huth:2021bsp}, an independent measurement of the Hubble constant~\cite{Abbott:2017xzu,Hotokezaka:2018dfi,Nakar:2020pyd,Dietrich:2020lps}, the confirmation that BNSs are a possible central engine for short gamma-ray bursts (GRBs)~\cite{GBM:2017lvd}, and the proof that BNSs are a production fabric for heavy elements~\cite{Abbott:2017wuw,Cowperthwaite:2017dyu,Smartt:2017fuw,Kasliwal:2017ngb,Kasen:2017sxr,Watson:2019xjv,Rosswog:2017sdn}. 
In contrast to GW170817, the second BNS detection GW190425~\cite{Abbott:2020uma} was not accompanied by any electromagnetic signature. Likely, this was caused by the large total mass of GW190425, which was larger than the mass of BNS systems previously observed in our galaxy~\cite{Farrow:2019xnc}. Given the unexpected large total mass of GW190425, it might also seem plausible that upcoming gravitational wave (GW) detections will originate from BNS systems with mass ratios that differ noticeably from the more common equal mass case. 

Theoretically, the expected NS mass ranges within $(1.0$ -- $ 2.3) M_\odot$. Bounds on the minimum mass of NSs come from the NS formation scenario (gravitational-collapse) and from observations of low mass NSs~\cite{Rawls:2011jw,Ozel:2012ax}, despite the fact that such measurements typically have large uncertainties. A lower bound on the maximum NS mass arises from Shapiro-time delay measurements of massive pulsars such as PSR~J0348+0432~\cite{Antoniadis:2013pzd}, PSR~J1614-2230~\cite{Arzoumanian:2017puf}, PSR~J0740+6620~\cite{Demorest:2010bx,Fonseca:2021wxt}. Contrary, the interpretation that the BNS merger GW170817 formed a BH~\cite{Margalit:2017dij,Ruiz:2017due,Rezzolla:2017aly,Shibata:2019ctb} provides an upper bound on the maximum NS mass; cf.~\cite{Dietrich:2020lps} for a more complete assessment of the NS maximum mass posterior distribution. Based on these considerations, the theoretical upper bound on the mass ratio of a BNS system is about  $q=M_1/M_2\lesssim 2.3$. 

Comparing this theoretical limit to the observed population of BNSs in our galaxy, one finds that none of the observed systems has such a large mass ratio. In fact, most BNS clusters are around the equal mass case~\cite{Lattimer:2019eez}, but there are discoveries about compact binary systems with mass ratio of $q \approx 1.3$, e.g., Refs.~\cite{Martinez:2015mya,Lazarus:2016hfu}. Hence, BNSs with larger mass ratios might exist and our current sample of observed BNS systems is simply too limited. Similarly, population synthesis studies predict a wider range of masses and mass ratios up to $q\approx 1.9$, e.g.,~\cite{Dominik:2012kk}. Given the rising field of GW astronomy and the uncertainty in the mass ratio of upcoming BNS detections, a careful investigation of GW signals from high-mass ratio systems seems necessary. \\

Overall, the analysis of GW signals requires accurate theoretical models that can be cross-correlated with the observational data. For BNS systems, the existing GW models can be classified into three categories: analytical Post-Newtonian (PN) models~\cite{Blanchet:2013haa}, semi-analytical waveforms based on the effective-one-body (EOB) approach~\cite{Hotokezaka:2015xka,Hinderer:2016eia,Steinhoff:2016rfi,Akcay:2018yyh,Nagar:2018plt,Bernuzzi:2014owa,Dietrich:2017feu,Nagar:2018gnk}, and phenomenological approximants~\cite{Kawaguchi:2018gvj,Dietrich:2017aum,Dietrich:2018uni,Dietrich:2019kaq}. To develop and validate these models, one requires numerical-relativity (NR) simulations for an accurate description of the merger process based on first principles. While there have been several studies addressing the effect of high mass ratios on the dynamics and merger process, e.g.,~\cite{Lehner:2016lxy,Dietrich:2016hky,Radice:2018pdn,Bernuzzi:2020txg,Papenfort:2022ywx}, there exists only a limited set of simulations usable for GW model development, i.e simulations using eccentricity-reduced initial data and with an accurate error estimate for the GW phase based on a clear convergence. In this article, we will overcome these limitations and produce a small set of highly accurate simulations. We simulated a total of four physical systems with mass ratios $q=1.25, 1.50, 1.75,$ and $2.00$ and a fixed total mass of $M=2.7 M_\odot$. We will use this new set of simulations to study the performance of existing GW models for such large mass ratios and will further investigate if techniques employed for modelling the higher-order modes in binary black hole models could also be used for the tidal contributions of the GW phase. 

Throughout this work we use geometric units, setting $c=G=M_\odot=1$, though we will sometimes include $M_\odot$ explicitly or quote values in CGS units for better understanding.

\section{Methods}
\label{sec:methods}

\subsection{Numerical setups}

\begin{table}[t]
\caption{Grid configurations. The columns refer to the resolution 
name, the number of levels $L$, the number of moving box levels 
$L_{\rm mv}$, the number of points in the non-moving boxes $n$, the 
number of points in the moving boxes $n_{\rm mv}$, the grid spacing in 
the finest level $h_6$ covering the neutron star, and the grid spacing in 
the coarsest level $h_0$. The 
grid spacing is given in units of 
$M_{\odot}$.} \label{tab:grid}
\begin{tabular}{ccccccc}
\toprule
Name & $L$ & $L_{\rm mv}$ & $n$ & $n_{\rm mv}$ & $h_6$ & $h_0$ \\ \hline
R1 & 7 & 4 & 128 & 64 & 0.249 & 15.936 \\
R2 & 7 & 4 & 192 & 96 & 0.166 & 10.624 \\
R3 & 7 & 4 & 256 & 128 & 0.1245 & 7.968 \\
R4 & 7 & 4 & 384 & 192 & 0.083 & 5.312 \\ 
R5 & 7 & 4 & 512 & 256 & 0.06225 & 3.984 \\ \hline 
\hline
\end{tabular}
\end{table}

Our numerical simulations are based on SGRID initial data~\cite{Tichy:2009yr,Tichy:2012rp,Dietrich:2015pxa,Tichy:2019ouu}. 
SGRID is a pseudospectral code that employs surface fitting coordinates to solve the Einstein Constraint Equations using the extended conformal thin sandwich formulation~\cite{York:1998hy}, and employs the constant rotational velocity approach to describe the NSs with arbitrary rotational profile~\cite{Tichy:2011gw,Tichy:2012rp}. 

The dynamical evolutions are performed with the BAM code~\cite{Brugmann:2008zz,Thierfelder:2011yi,Dietrich:2015iva,Bernuzzi:2016pie,Dietrich:2018phi}. BAM uses the Z4c formulation~\cite{Bernuzzi:2009ex,Hilditch:2012fp} of the Einstein equations along with 1+log and gamma-driver shift conditions~\cite{Bona:1994a,Alcubierre:2002kk,vanMeter:2006vi} for the gauge evolution. Matter variables are evolved using the $3+1$-conservative Eulerian formulation of general-relativistic hydrodynamics (GRHD). The system of equations is closed by an equation of state (EOS) that is a piecewise-polytropic fit for the SLy~\cite{Read:2008iy} EOS, which is in broad agreement with recent multi-messenger constraints, with an additional thermal contribution to the pressure given by $p_{\rm th} = (\Gamma_{\rm th} - 1) \rho$, where we set $\Gamma_{\rm th} = 1.75$~\cite{Bauswein:2010dn}.

BAM's numerical domain is divided into a hierarchy of cell centered nested Cartesian grids consisting of $L$ levels labeled by $l=0,...,L-1$. Each level $l$ contains one or more Cartesian boxes with constant grid spacing $h_l$ and $n$ (or $n^{\rm mv}$) number of points per direction. The resolution in each level is given as $h_l = h_0/2^l$. Higher levels $l \geq l_{\rm mv}$ move dynamically according to the technique of `moving boxes’ and follow the motion of the neutron stars. An overview about the grid configuration for different resolutions is given in Tab.~\ref{tab:grid}, where the outer boundary of the computational domain is set at radius $R_0 \approx 1020M_\odot$.

\subsection{Configurations} 

\begin{table}[t]
\caption{Properties of the individual stars used for our BNS simulations.  
The first column gives the configuration name, the second column the employed mass ratio $q=M^A/M^B\geq 1$, the next four column give the gravitational masses of the individual stars $M^{A,B}$ and the baryonic masses of the individual stars $M _b ^{A,B}$. \label{tab:config_single}} 
\begin{tabular}{cccccc}
\toprule Name & $q$ & $M^A$ & $M^B$ & $M^A_b$ & $M^B_b$ \\ \hline
\sone & 1.25 & 1.5001 & 1.2001 & 1.6834 & 1.3117 \\ 
\stwo & 1.50 & 1.6201 & 1.0801 & 1.8388 & 1.1688 \\
\sthree & 1.75 & 1.7183 & 0.9819 & 1.9695 & 1.0542 \\
\sfour & 2.00 & 1.8002 & 0.9001 & 2.0813 & 0.9601 \\ \hline \hline
\end{tabular}
\end{table}

\begin{table}[t]
\caption{Properties of our BNS simulations.  
The columns give the configuration name, the residual eccentricity $e$, 
the initial GW frequency $M\omega^0 _{2,2}$ of the (2,2)-mode, 
the Arnowitt-Deser-Misner (ADM) mass $M_\text{ADM}$, 
and the angular momentum $J_\text{ADM}$. 
All configurations were evolved with the resolutions of 
Tab.~\ref{tab:grid}. \label{tab:config_double}.} 
\begin{tabular}{ccccc}
 Name & $e$ [$10^{-4}$] & $M\omega^0_{2,2}$ [$10^{-2}$] & $M_\text{ADM}$ & $J_\text{ADM}$ \\ \hline \hline
\sone  & 1.6413 & 3.2241 & 2.6804 & 7.9252 \\
\stwo  & 3.0940 & 3.2231 & 2.6809 & 7.7035 \\
\sthree  & 3.1148 & 3.2220 & 2.6816 & 7.4280 \\
\sfour & 3.4273 & 3.2213 & 2.6824 & 7.1353 \\ \hline \hline
\end{tabular}
\end{table}

For this work, we prepare a set of four different physical BNSs systems. Each setup has the same total gravitational mass in isolation but different mass ratio, see Tab.~\ref{tab:config_single}. In order to eliminate eccentric contributions of the orbit to the gravitational wave during the inspiral, we perform between three and four steps of the eccentricity reduction procedure~\cite{Moldenhauer:2014yaa,Dietrich:2015pxa} obtaining values lower than $3.5\times 10^{-4}$. All configurations start at the same initial frequency and the number of orbits before merger is approximately sixteen. See Tab.~\ref{tab:config_double} for important parameters characterizing the initial properties of the binaries.

\section{Results}
\label{sec:results}

\subsection{Gravitational waves}

We extract GWs from our simulations using the Newman-Penrose formalism~\cite{Newman:1962} 
based on the curvature scalar $\Psi_4$. The GW strain is then computed from $\Psi_4 = \ddot{h}$, where we use the frequency domain integration outlined 
in Ref.~\cite{Reisswig:2010di}. 
For our purpose, it is convenient to decompose $\Psi_4$ as well as $h$
into individual modes by employing spherical harmonics with spin weight -2, i.e., 
\begin{equation}
    h = \sum_{l=2}^{l_{\rm max}}\  \sum_{m=-l}^{m\leq l}\ ^{-2}Y_{l,m} h_{l,m}.
\end{equation}

\begin{figure}
\centering
\includegraphics[width=0.48\textwidth]{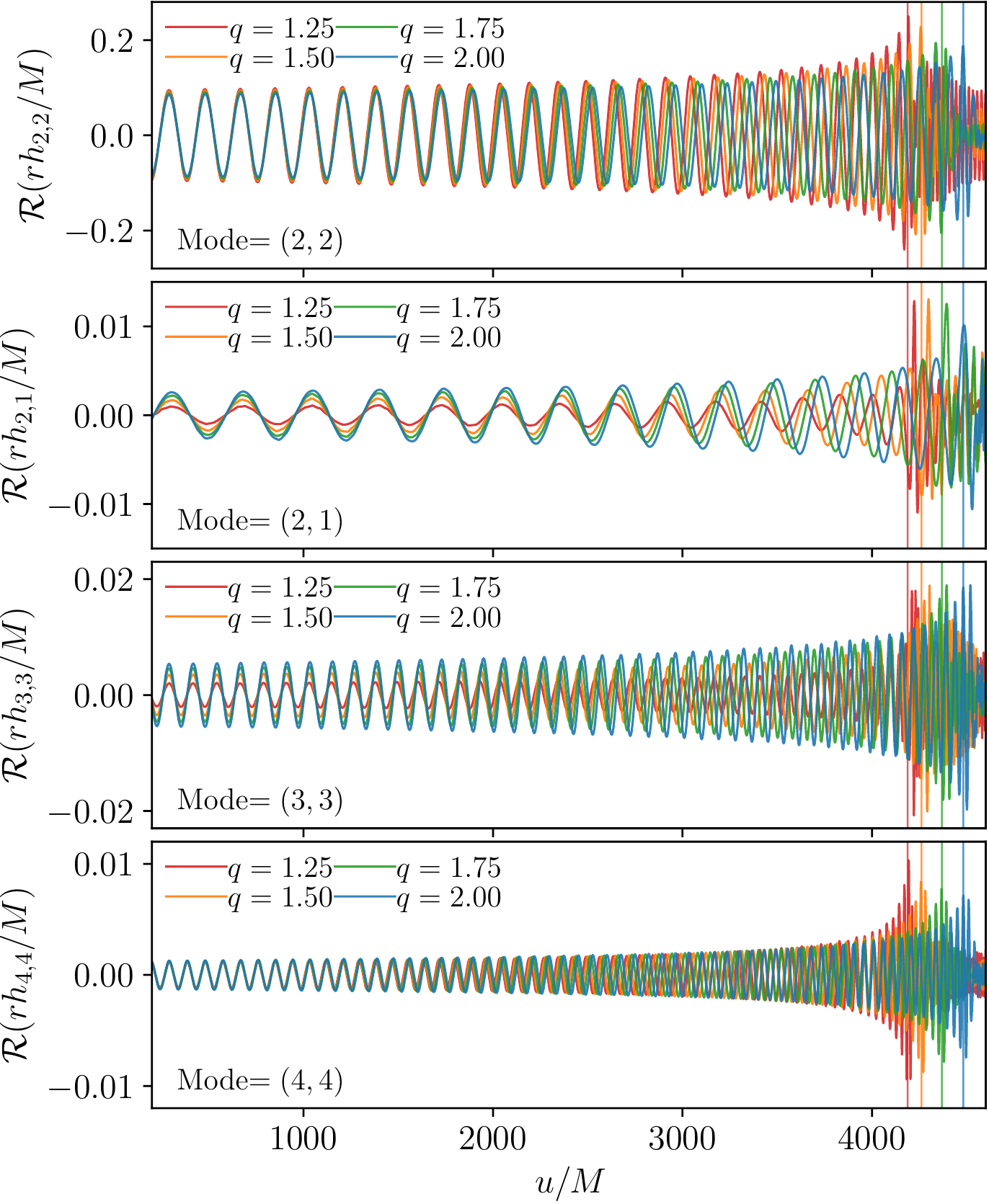}
\caption{Real part for different modes of the GW $rh$ as a function of the retarded time $u/M$, where $r$ is the coordinate radius. As expected, the (2,2)-mode is the dominant mode and the amplitudes of the (2,1)-mode and (3,3)-mode increase with increasing mass ratio. The vertical lines in the plot indicate the merger time for resolution R5, from which we note that systems with higher mass ratios starting at the same frequency merge later.}
\label{fig:waves_qualitative}
\end{figure}

In Fig.~\ref{fig:waves_qualitative}, 
we show the dominant (2,2)-mode of 
the GW signal in the top panel and the subdominant (2,1), (3,3), (4,4) 
modes in the following rows. 
For completeness, we want to summarize some of the main findings, which, however, have already been found by previous studies and also analytical computations, e.g.,~\cite{Blanchet:2013haa}. 

(i) The (2,2)-mode is dominanting and has the largest amplitude. 

(ii) The amplitude of all ($l,m)$-modes with odd $m$ values increases with increasing mass-ratio. In the equal mass case, due to the symmetry of the system, these modes are zero (during the inspiral). 

(iii) The frequencies of the individual modes depend on the value of $|m|$ and scale roughly according to $\omega_{2,2}\cdot |m/2|$.

(iv) For non-precessing systems the $m$ and $-m$ modes are identical (not shown in Fig.~\ref{fig:waves_qualitative}).

\subsection{Convergence properties}

\begin{figure*}
\centering
\includegraphics[width=0.96\textwidth]{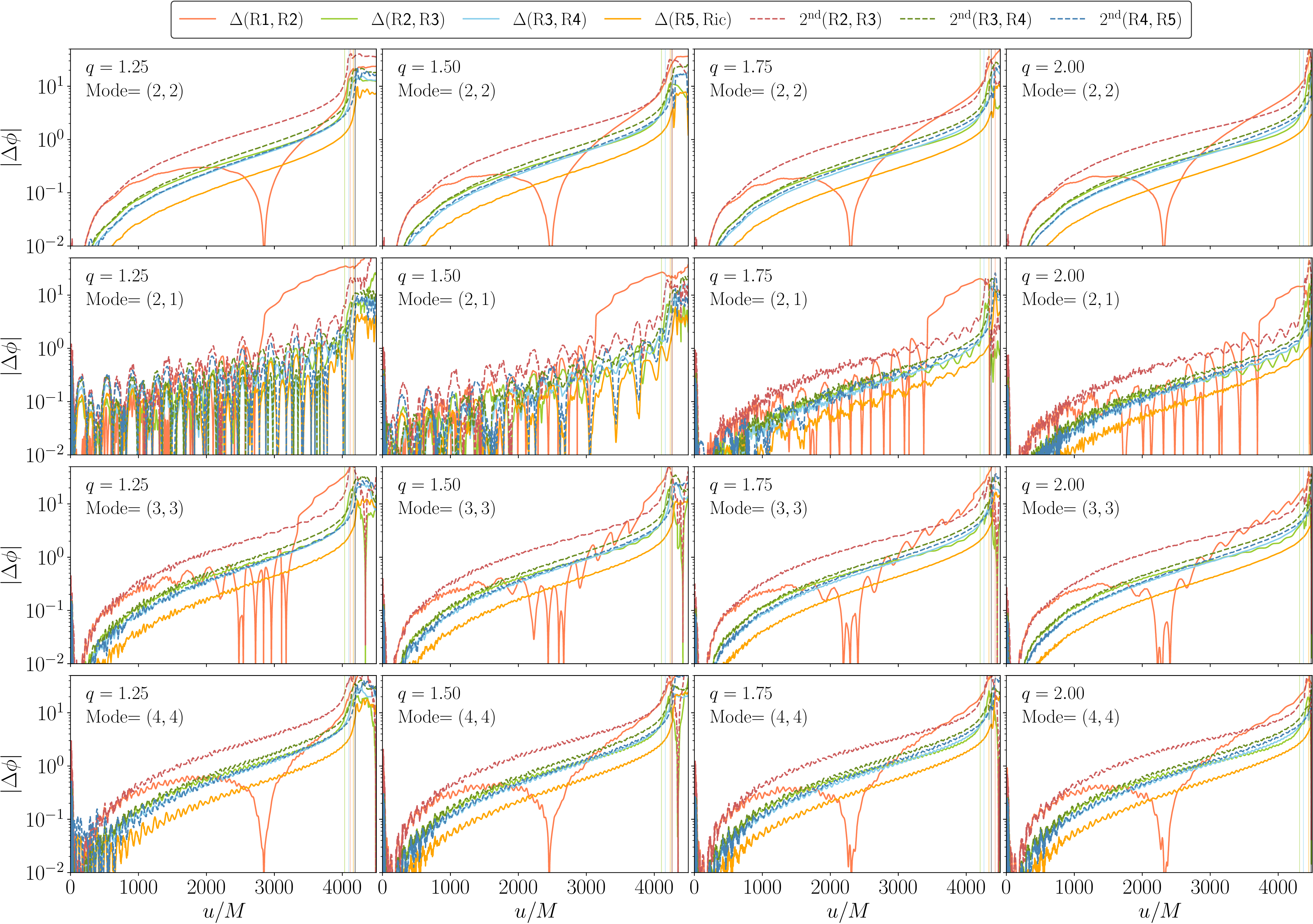}
\caption{Convergence for all important modes. We show the phase difference between different resolutions (solid lines) and the rescaled phase difference assuming second order convergence (dashed). 
Individual rows refer to the (2,2), (2,1), (3,3), (4,4) mode (from top to bottom), the individual columns refer to the setups with increasing mass ratio. The vertical lines in the plot indicate the merger time.}
\label{fig:convergence_all}
\end{figure*}

To investigate the convergence properties of our simulations, we write the GWs as
\begin{equation}
    h_{l,m} = A_{l,m} e^{-\i \phi_{l,m}}
\end{equation}
with the amplitude $A_{l,m}$ and the phase $\phi_{l,m}$. Given that GW-astronomy relies mainly on a proper estimate of the GW phase, we will focus our discussion on the convergence properties for the GW phase. For this purpose, we show the convergence of all modes (for all setups and resolution) in Fig.~\ref{fig:convergence_all}. Based on our previous studies~\cite{Bernuzzi:2016pie,Dietrich:2018upm}, we expect to obtain second order convergence for our numerical simulations with respect to the (2,2)-mode. 

In this work, we tested that this convergence order is also present in the higher-order modes. For this purpose, we scale the phase difference between two different resolutions under the assumption of second-order convergence and find good agreement for the high-resolution simulations. 
Let us outline key features of the plot:

(i) Phase differences with respect to the lowest resolution R1 indicate that for such a resolution the simulations do not reach the convergent regime. Hence, a smaller grid spacing, i.e., a higher resolution, is required.

(ii) Independent of the mode that we consider, an increasing mass ratio leads to a slight reduction of the  convergence order. 

(iii) Due to the very small amplitude, the assessment of the convergence properties for the $(2,1)$-mode is problematic for mass ratios below the $q=1.75$ (cf.~Fig.~\ref{fig:waves_qualitative}).

In addition to the phase difference between individual resolutions, we also show the difference between the GW phase for the highest resolution and Richardson-extrapolated waveforms for which we assumed second order convergence during the extrapolation, see Refs.~\cite{Bernuzzi:2016pie,Dietrich:2018upm} for more details about this procedure. 

\subsection{Tidal contribution in higher modes}

\begin{figure}[t]
\centering
\includegraphics[width=0.48\textwidth]{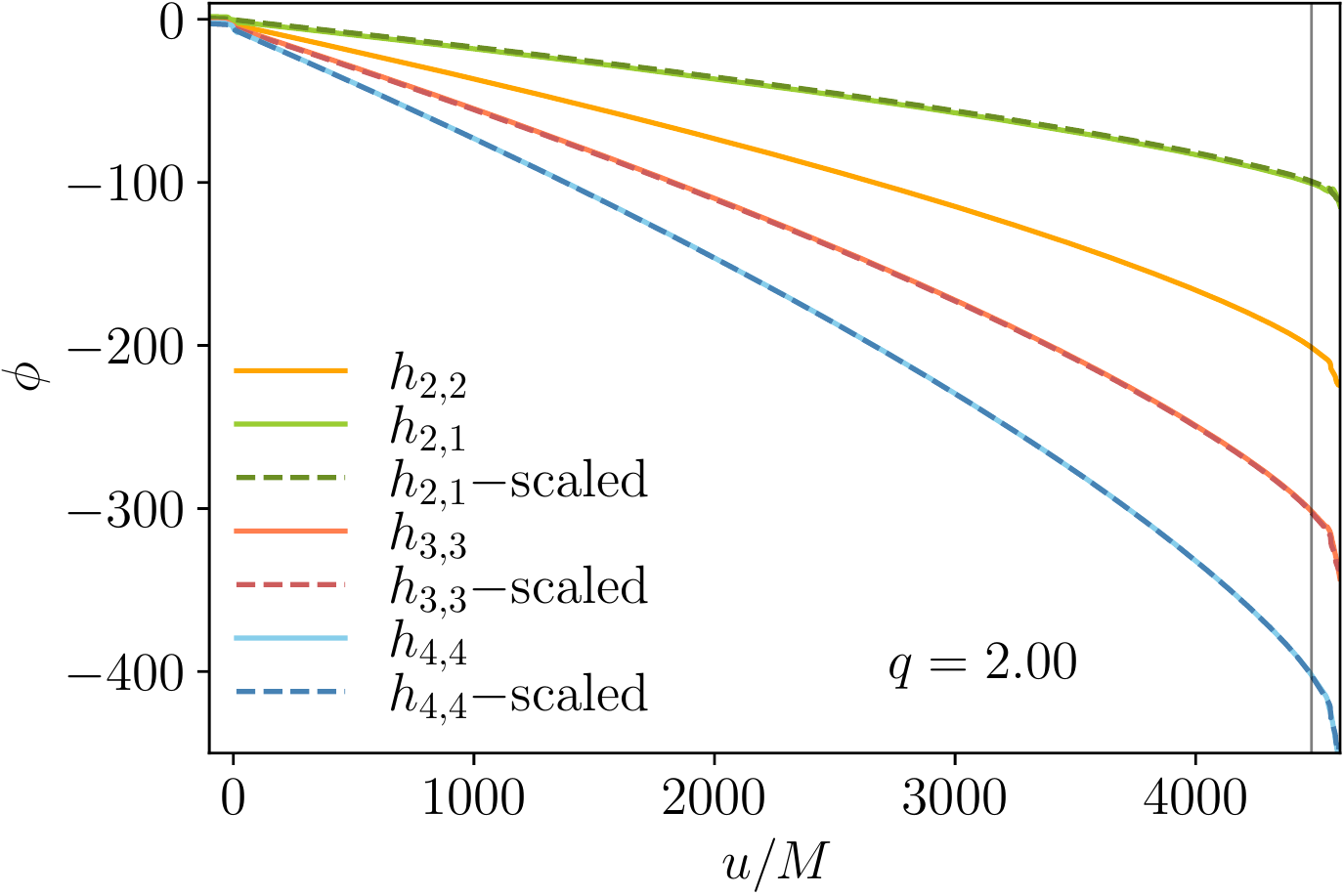}
\caption{Accumulated phase of our simulations for the $q=2.00$ setup, where we show the individual modes with different colors and rescale the dominant (2,2)-mode according to the description given in the text to mirror the evolution of the subdominant modes. The scaling of the accumulated phase is verified for all mass ratios.}
\label{fig:scalingNS}
\end{figure}

Several GW models that model higher-order modes start from a description of the dominant (2,2)-mode and then rescale this mode to obtain subdominant mode 
predictions, e.g.,~\cite{London:2017bcn,Khan:2019kot,Cotesta:2020qhw}. 
The scaling relations are 
\begin{equation}
\phi_{l,m} = m \phi_{\rm orbital} + \Delta \phi_{l,m} \label{eq:scaling_relation}
\end{equation}
with the orbital phase $\phi_{\rm orbital}$ and 
\begin{equation}
\Delta \phi_{2,2} \rightarrow 0,
\Delta \phi_{2,1} \rightarrow \pi/2,
\Delta \phi_{3,3} \rightarrow -\pi/2,
\Delta \phi_{4,4} \rightarrow \pi. \label{eq:scaling_relation_coefficients}
\end{equation}

Under the assumption of $2\phi_{\rm orbital} \approx \phi_{2,2}$, 
this allows us to use information from the dominant (2,2)-mode to predict 
the evolution of the subdominant modes. 
As expected, this scaling relation also clearly applies for our 
NR simulations using the highest resolution data. 
This becomes visible in Fig.~\ref{fig:scalingNS}, where we rescale 
the (2,2)-mode contribution according to Eqs.~\eqref{eq:scaling_relation}
and \eqref{eq:scaling_relation_coefficients} and find overall perfect agreement between 
our data and the analytical predictions.

\begin{figure}[t]
\centering
\includegraphics[width=0.48\textwidth]{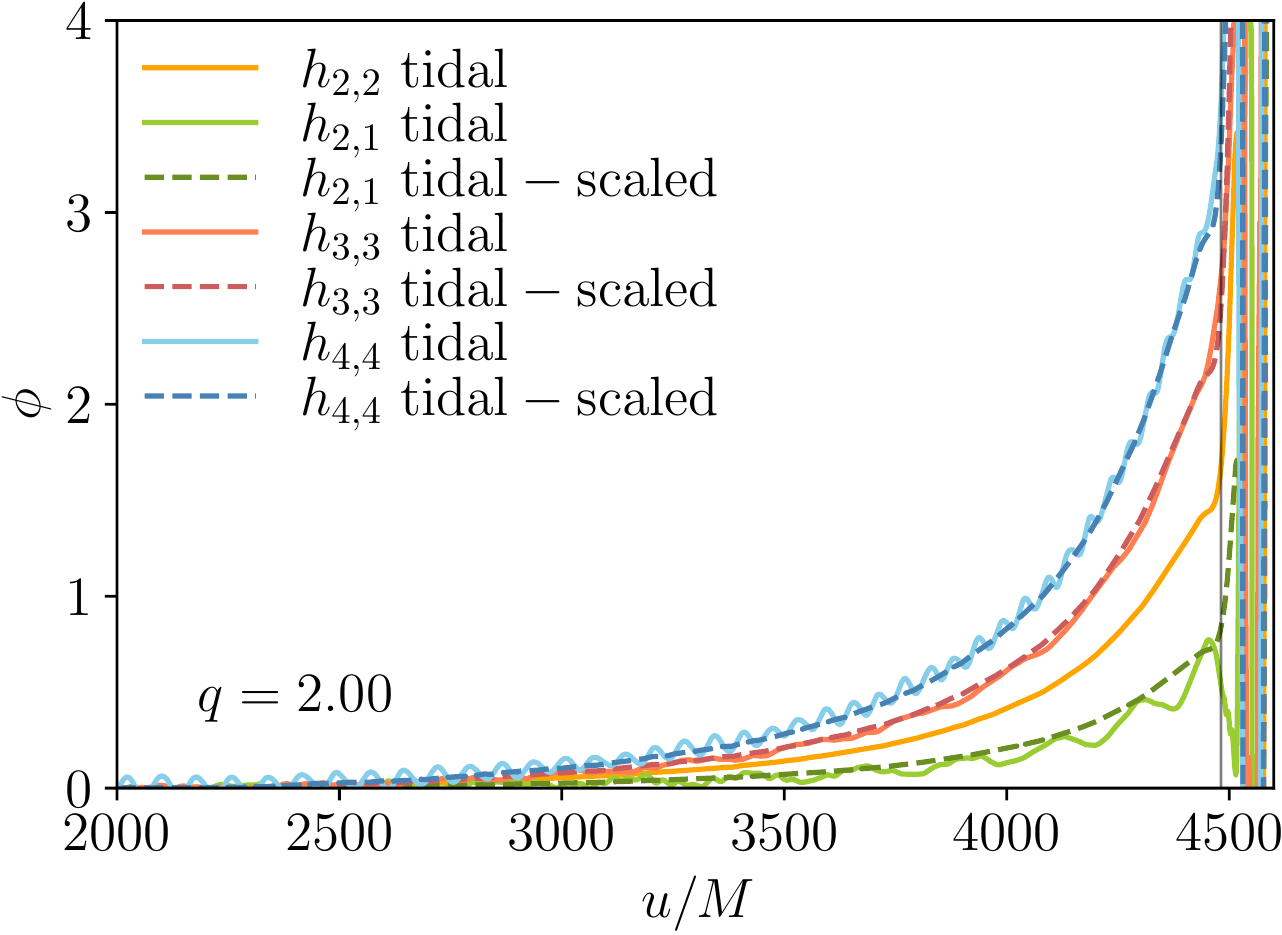}
\caption{Rescaling of the tidal contribution $\phi^{\rm Tidal}_{l,m}$ for different modes similar to Fig.~\ref{fig:scalingNS} for the entire phase contribution of each mode. We verify that the scaling of the tidal contribution is valid for all mass ratios.}
\label{fig:scalingTidal}
\end{figure}

In addition to the investigation of the entire mode content, we also want to investigate 
if the rescaling can be applied to the individual contributions, see Eq. (\ref{eq:contributions}). Overall, also for the individual contributions our NR simulations can serve as a validation set and they prove that BNS simulations can reach an accuracy level were a reliable modelling of subdominant modes become possible.\\ 

An approach that is intensively used during the modelling of BNS and BHNS systems is the assumption that 
the GW phase can be decomposed into different components, keeping for simplicity the (2,2)-mode, 
we get 
\begin{equation}
\phi_{2,2}=\phi_{2,2}^{\rm BBH}+\phi_{2,2}^{\rm SO}+\phi_{2,2}^{\rm SS}+\phi_{2,2}^{\rm Tidal} +...
\label{eq:contributions}
\end{equation}
with the non-spinning BBH contribution $\phi_{2,2}^{\rm BBH}$, the spin-orbit contributions $\phi_{2,2}^{\rm SO}$, the spin-spin contribution $\phi_{2,2}^{\rm SS}$, and the tidal contributions $\phi_{2,2}^{\rm Tidal}$. 
In this article, we will particularly focus on $\phi_{2,2}^{\rm Tidal}$, as well as higher-order contributions $\phi_{2,1}^{\rm Tidal}$, $\phi_{3,3}^{\rm Tidal}$, and $\phi_{4,4}^{\rm Tidal}$, the spin contributions are zero in our spinless configurations. For this purpose, we substract from our BNS NR waveforms the BBH contributions computed with the \texttt{SEOBNRv4T}~\cite{Hinderer:2016eia} model (in its form currently implemented in LALSuite~\cite{lalsuite}).  

Fig.~\ref{fig:scalingTidal} summarizes our findings: The scaling relations, Eq.~\eqref{eq:scaling_relation}, also apply  for the tidal contribution. This is of special importance for the possibility to extend existing BNS or BHNS models that currently purely model the dominant $(2,2)$-mode, e.g., \cite{Dietrich:2018uni,Dietrich:2019kaq,Matas:2020wab,Thompson:2020nei}, to model 
the tidal contributions present in higher modes.

\section{Comparison with existing GW models}
\label{sec:comparison}

\subsection{Model validation}

\begin{figure*}
\centering
\includegraphics[width=0.96\textwidth]{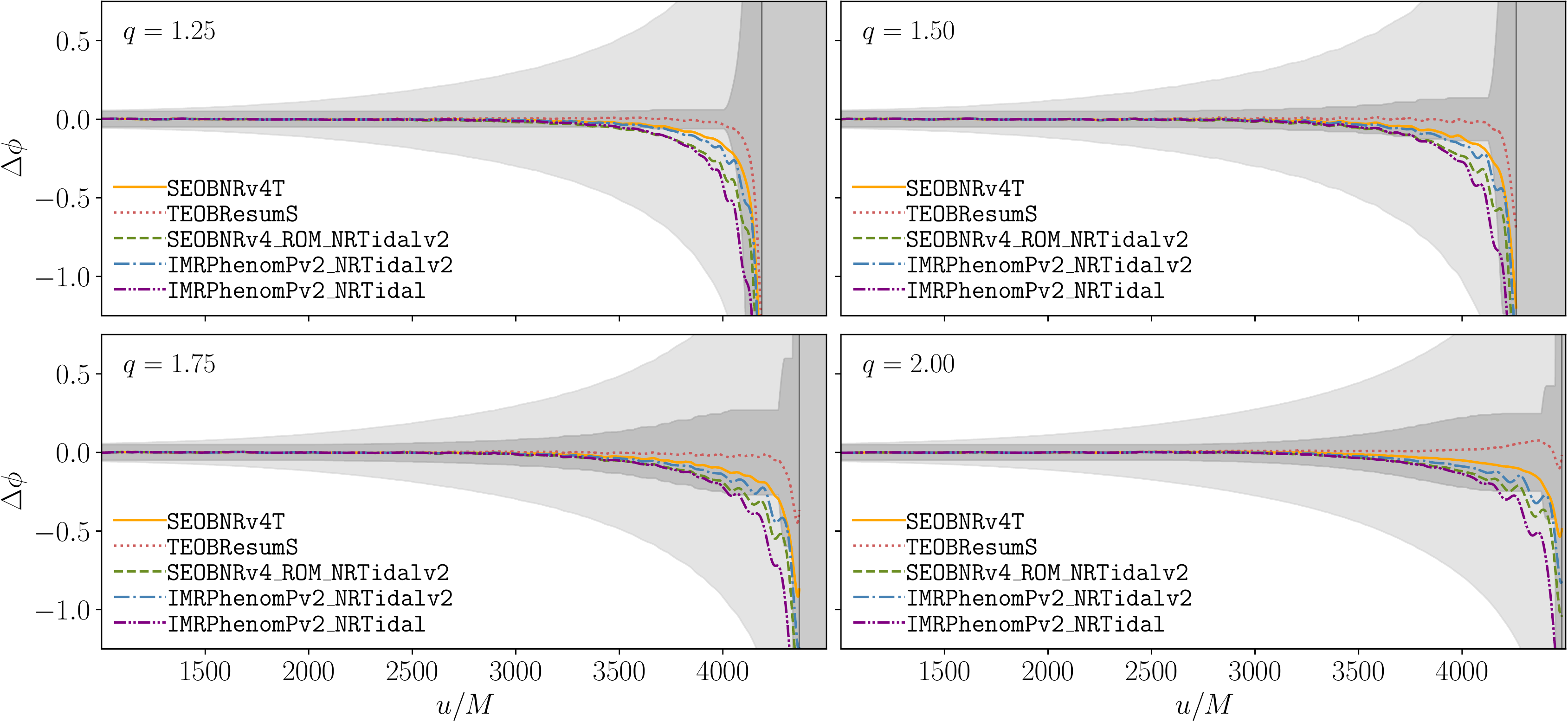}
\caption{Phase difference of the (2,2)-mode between the Richardson-extrapolated data and the different GW models for all values of $q$. The alignment of their phases have been performed on an early stage in the inspiral before subtracting. All GW models fall within our more conservative light gray error band.} 
\label{fig:comparison_all}
\end{figure*}

An important application for our new NR simulations is the possibility to quantify the performance of GW models currently employed in GW analysis. For this purpose, 
we compare our simulation results with five different state-of-the-art GW models, namely, 
\texttt{SEOBNRv4T}~\cite{Hinderer:2016eia}, \texttt{TEOBResumS}~\cite{,Nagar:2018plt}, \texttt{SEOBNRv4\_ROM\_NRTidalv2}~\cite{Dietrich:2019kaq}, \texttt{IMRPhenomPv2\_NRTidal}~\cite{Dietrich:2018uni}, and \texttt{IMRPhenomPv2\_NRTidalv2}~\cite{Dietrich:2019kaq}. For completeness, we summarize key features of each model in the appendix. 

In Fig.~\ref{fig:comparison_all}, we show the phase difference between the GW obtained from different GW models and the Richardson-extrapolated signal. 
Before subtracting, we first align the model waveform with respect to the Richardson-extrapolated data by minimizing their phase difference on an early stage of the inspiral, where $M\omega_{2,2} \in [0.035,0.040]$. For comparison, we also show in the figure two different error bands. The conservative light gray error band ($\pm \epsilon$) is estimated using two terms, i.e. $\epsilon^2 = \epsilon^2_{\rm Ric} + \epsilon^2_{\rm Ext}$. The first term is obtained through the difference between the Richardson-extrapolated value and the NR simulation with the highest resolution, i.e. $\epsilon_{\rm Ric}$ = $\Delta({\rm Ric} - {\rm R5})$. The second term, $\epsilon_{\rm Ext}$, is obtained calculating the phase difference between two different extraction radii. In our case we consider radii $700M_\odot$ and $900M_\odot$ in the highest resolution available. This last term is almost constant and is the dominant error in $\epsilon$ until approximately $2000M_\odot$. The dark gray error band ($\pm \epsilon_\Delta$) is an error for the third order Richardson extrapolation itself, and it is given by $\epsilon^2_\Delta = {\rm max}[\epsilon^2_{\Delta{\rm Ric}} + \epsilon^2_{\rm Ext}]$, where max indicates that we are keeping the highest value as time passes by, to ensure a monotonically increasing uncertainty. The first term, $\epsilon_{\Delta{\rm Ric}}$, is the difference between two different Richardson extrapolations. One extrapolation is done with resolutions R3 and R4, ${\rm Ric}_{34}$, and the other one with resolutions R4 and R5, ${\rm Ric}_{45}$. Thus, this term can be written as $\epsilon_{\Delta{\rm Ric}} = \Delta({\rm Ric}_{45}-{\rm Ric}_{34})$. The second term is the same present in $\epsilon$. In this case, $\epsilon_{\Delta{\rm Ric}}$ becomes larger than $\epsilon_{\rm Ext}$ only in the final stage of the inspiral.

Independent of the mass ratio, we find that the phase difference between our Richardson-extrapolated waveforms and the waveform approximants are in good agreement. All models fall within our more conservative error measure $\epsilon$ and therefore can not be clearly discarded/disfavored in any way. 
Nevertheless, we do find interesting patterns in our analysis. 
Noticeably, the \texttt{IMRPhenomPv2\_NRTidal} model shows the largest difference to our NR-based data. The two NRTidalv2 models  \texttt{IMRPhenomPv2\_NRTidalv2} and  \texttt{SEOBNRv4\_ROM\_NRTidalv2} perform slightly better. However, all models fall shortly before merger out of our tighter error band $\epsilon_\Delta$; cf.\ dark gray region. 
The two time domain EOB models \texttt{SEOBNRv4T} and \texttt{TEOBResumS} perform best with, in particular, \texttt{TEOBResumS} always staying within our tight error band. 
Interestingly, in general, we find that the phase difference between our prediction and the GW model predictions is negative, i.e., Richardson-extrapolated values (3rd. order) minus \texttt{Model}. The only exception from this `rule' is the simulation for the highest mass ratio, where \texttt{TEOBResumS} shows a positive phase difference with respect to our NR data close to merger. While one could speculate that this is caused by an overestimation of tidal effects for \texttt{TEOBResums} for large mass ratios, it is more likely that this is caused by the fact that the Richardson-extrapolated result is less accurate since the convergence order reduces slightly before the merger, as seen in Fig.~\ref{fig:convergence_all}. 

\begin{figure}[t]
\centering
\includegraphics[width=0.48\textwidth]{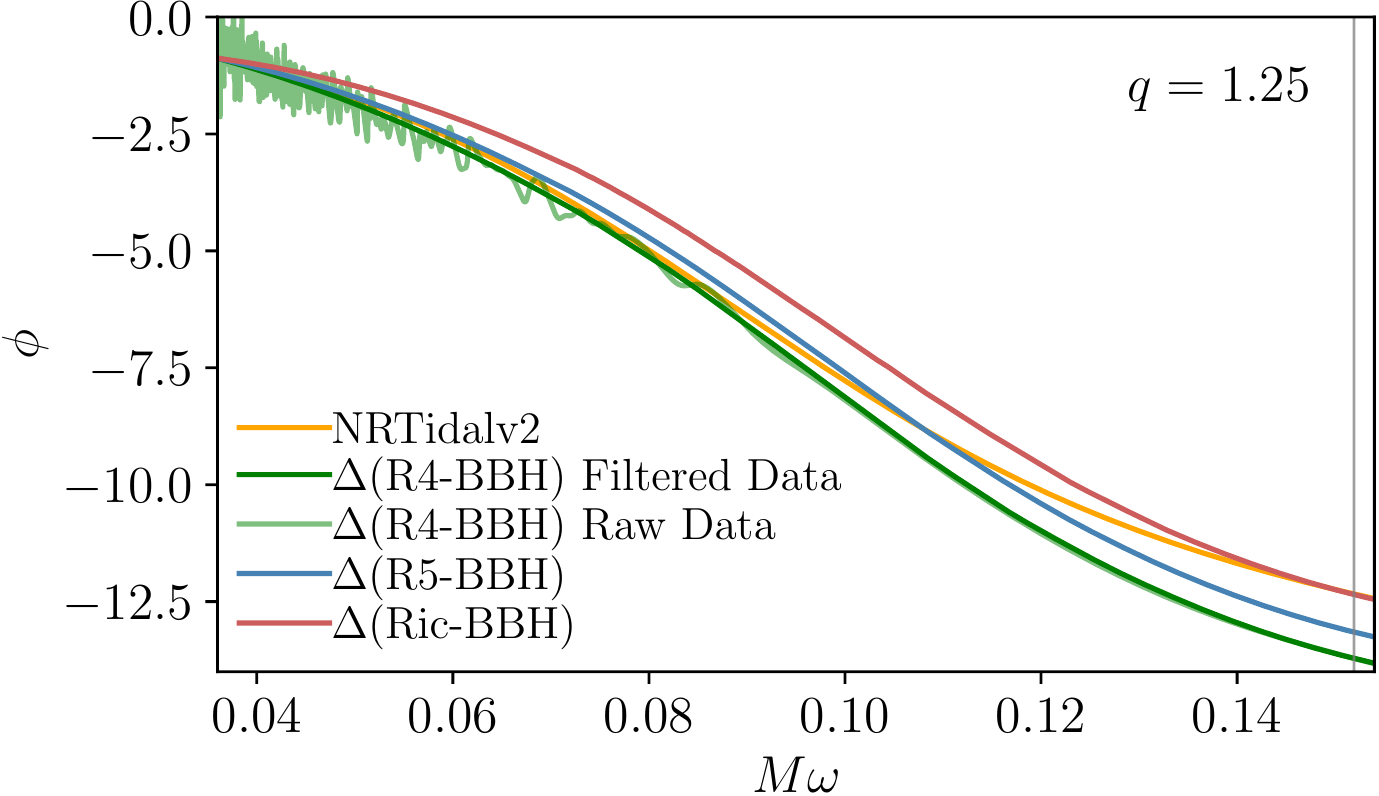}
\caption{Tidal contributions of the (2,2)-mode of our NR simulations and the Richardson-extrapolated data for the case $q=1.25$. For comparison we align them with the NRTidalv2 tidal model. To show the effect of the filtering procedure on the results, we present the R4 configuration with (dark green) and without (light green) the Savitzky-Golay filtering.}
\label{fig:phivsMomq125}
\end{figure}

\subsection{Outlooks to improve the NRTidal model}

\begin{figure*}[t]
\centering
\includegraphics[width=\textwidth]{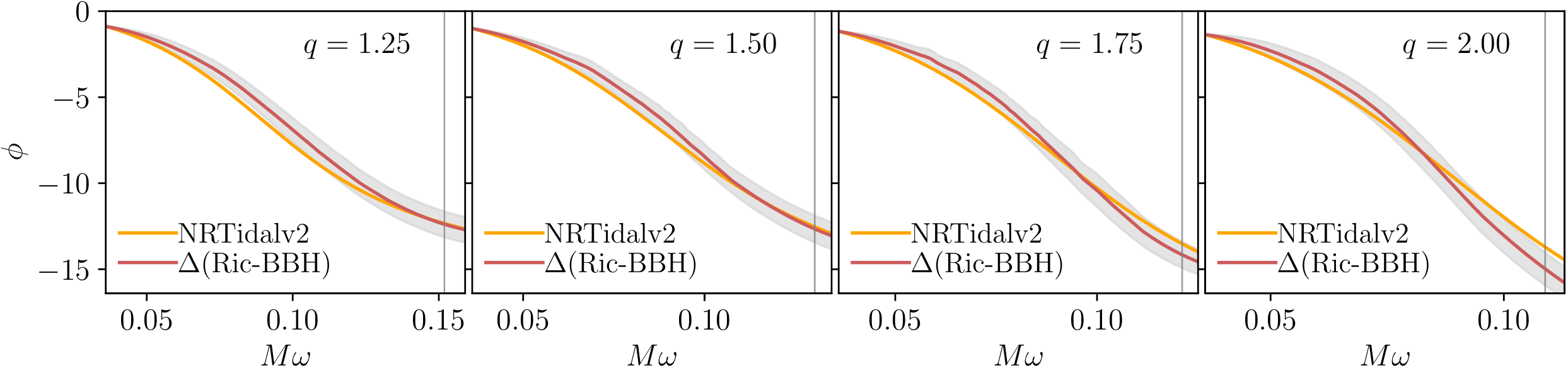}
\caption{Comparison of the NRTidalv2 model (orange) with the tidal contribution of the (2,2)-mode of the Richardson-extrapolated data (brown) for all values of $q$. In general, the model has some deviations  but behaves correctly for all mass ratios and falls within our error band during most of the frequency interval.}
\label{fig:phivsMomall}
\end{figure*}

In addition to the general model comparison presented in the previous subsection, we want to focus on the potential to further improve (and test) the NRTidal description~\cite{Dietrich:2017aum}. 
In general, the NRTidal phase contribution, used to augment existing BBH models, is based on the possibility to extract $\phi(\omega)$. So far only the (2,2)-mode has been used for any NRTidal-based model, but based on Fig.~\ref{fig:scalingTidal}, also higher modes could be modeled. Hereafter we name for simplicity $\omega_{2,2}$ by $\omega$.

Another important point is that, while the NRTidalv2 model incorporated next-to-leading order mass ratio effects, the model has been calibrated purely to equal-mass simulations; see~\cite{Dietrich:2019kaq}.
To check the robustness of this approach, we first extract the tidal contribution to $\phi_{2,2}(\omega)$ from our simulations.

To obtain the tidal contribution $\phi^{\rm Tidal}_{2,2}(\omega)$ from the NR simulation, we need to subtract the BBH contribution $\phi^{\rm BBH}_{2,2}(\omega)$ from $\phi_{2,2}(\omega)$. As a first step, we use again \texttt{SEOBNRv4T} to model $\phi(t)$ for a BBH system with the same component masses. Then, we can calculate $\omega = \partial_t \phi$ and obtain $\phi(\omega)$. Afterwards, we align the BBH phase and the BNS phase at a given frequency, where we use here $M\omega=0.036$ as the reference frequency for alignment. Finally, after obtaining the tidal contribution, we perform a further alignment with the NRTidalv2 model using the same reference frequency. The tidal contributions $\phi^{\rm Tidal}_{2,2}(\omega)$ found in this way have an oscillatory behavior for low frequencies. To reduce these oscillations, we employ the Savitzky–Golay filtering. In Fig.~\ref{fig:phivsMomq125} we present, for the $q=1.25$ case, the tidal contributions $\phi^{\rm Tidal}_{2,2}(\omega)$ for different NR resolutions and for the Richardson-extrapolated waveform. In dark and light green we show an example of the use of the Savitzky-Golay filtering, the original tidal contribution for resolution R4 (light green) has an oscillatory behavior near $M\omega = 0.04$ which is corrected after using the filtering (dark green). For resolutions R5 and Ric we plot only the filtered version.

In Fig.~\ref{fig:phivsMomall}, we compare the tidal contribution of the Richardson-extrapolated waveform, $\phi^{\rm Tidal}_{2,2~{\rm Ric}}(\omega)$, with the NRTidalv2 model for different mass ratio. Each tidal contribution is obtained following the procedure described in the above paragraph. The error band in these plots is calculated using the difference between the Richardson-extrapolated value and the highest NR resolution, i.e., $\epsilon^{\rm Tidal} = \pm \Delta[\phi^{\rm Tidal}_{2,2~{\rm Ric}}(\omega) - \phi^{\rm Tidal}_{2,2~{\rm R5}}(\omega)]$. Taking into account this conservative error, we see
that in general, the NRTidalv2 model behaves correctly for all mass ratios and falls within our error band during most of the frequency interval. But, there are deviations. These are most noticeable for $q=1.25$ (in the frequency $M\omega \approx 0.08$) and for $q=2.00$ (at frequencies $M\omega > 0.08$). Because of this, we still see room for improvements of the NRTidal phase description and we hope that our new set of NR data will be helpful towards future developments.

\section{Conclusion and Outlook}
\label{sec:conclusion}

In this article, we have performed simulations of four different BNS systems for mass ratios $q=1.25,1.50,1.75,2.00$. All physical configurations have been simulated with five different resolutions to allow for a proper error assessment. We find that the (2,2), (2,1), (3,3), (4,4) modes show 2nd order convergence, where for larger mass ratios the convergence order drops slightly around the moment of merger. We use our simulations to compute Richardson-extrapolated GW data that we compare against existing, state-of-the-art GW models. We find overall good agreement (within our estimated error bands), but a preference for the models \texttt{TEOBResumS} and \texttt{SEOBNRv4T}. 

Our simulations allow us also to verify that the estimated tidal contribution to the dominant (2,2)-mode can be rescaled to mimic tidal contributions for the subdominant modes. This is of special interest for the development of future BNS and BHNS models that will be required for the upcoming observing runs of the advanced GW-detector networks.

Finally, we verify that despite the calibration to equal-mass systems, the existing NRTidalv2 model is able to describe also high-mass ratio systems as presented here. Nevertheless, we expect that based on this set of simulation and upcoming simulations that are ongoing, there will be the possibility to further improve the NRTidal model to ensure a more reliable interpretation of future GW signals.

\begin{acknowledgments}
M.U. acknowledges support through the Coordena\c{c}\~ao de Aperfei\c{c}oamento de Pessoal de N\'ivel Superior - Brasil (CAPES) - Process number: 88887.571346/2020-00. H.G. acknowledges support through grant 2019/26287-0, S\~ao Paulo Research Foundation (FAPESP). W.T. was supported by the National Science Foundation under grants PHY-1707227 and PHY-2011729.

Computations have been carried out on the Dutch national e-infrastructure with the support of SURF Cooperative, project number 2019.021. 
Furthermore, parts of the simulations were performed on the national supercomputer HPE Apollo Hawk at the High Performance Computing Center Stuttgart (HLRS) under the grant number [project GWanalysis 44189].
\end{acknowledgments}

\appendix

\section{Summary of employed GW models}

\texttt{IMRPhenomPv2\_NRTidal} is based on the precessing BBH model \texttt{IMRPhenomPv2}~\cite{Hannam:2013oca, Khan:2015jqa} and uses the original NRTidal correction~\cite{Dietrich:2017aum,Dietrich:2018uni} to account for tidal contributions. In addition, 2PN and 3PN EOS-dependent spin-spin effects are employed to augment the existing BBH-baseline model. No further tidal amplitude corrections are employed. 

\texttt{IMRPhenomPv2\_NRTidalv2} is an update of \texttt{IMRPhenomPv2\_NRTidal}. In addition to \texttt{IMRPhenomPv2\_NRTidal}, this model uses the updated NRTidalv2~\cite{Dietrich:2018uni} description for the tidal contribution. It also employs a tidal amplitude correction, and it includes up to 3.5PN EOS-dependent effects in the spin-spin and cubic-in-spin contributions (including octupole-dependent terms).

\texttt{SEOBNR\_ROM\_NRTidalv2} is a frequency domain model based on the surrogate \texttt{SEOBNRv4\_ROM}~\cite{Bohe:2016gbl,Lackey:2018zvw}. The BBH model is augmented with the NRTidalv2 phase corrections~\cite{Dietrich:2018uni}, uses tidal amplitude corrections, and EOS-dependent spin-spin and cubic-in-spin corrections up to 3.5PN.

\texttt{SEOBNRv4T} is a time domain EOB 
model~\cite{Hinderer:2016eia,Steinhoff:2016rfi} based on the time domain BBH model~\cite{Bohe:2016gbl}. The model includes quadrupolar and octopolar dynamical tides, as well as EOS-dependent spin-induced quadrupole moment effects. In our work, we rely on the publicly available version of LALSuite~\cite{Hinderer:2016eia} that does not include the spin-dependence of the dynamical tidal effects as presented in~\cite{Steinhoff:2021dsn}.

\texttt{TEOBResumS} is a time domain EOB model. In contrast to \texttt{SEOBNRv4T}, \texttt{TEOBResumS} uses a gravitational self-force inspired expressions for the attractive tidal potential~\cite{Bernuzzi:2014owa} but restrict to adiabatic tidal effects. As for the other models, we rely on the publicly available \texttt{TEOBResumS} model that is part of LALSuite~\cite{Nagar:2018plt}. 


\bibliography{bns_high_q_eccred.bib}

\end{document}